\documentstyle[amscd,amssymb,12pt]{amsart} 
\input xypic \xyoption{curve}

\def\hcorrection#1{\advance\hoffset by #1 }
\def\vcorrection#1{\advance\voffset by #1 }

\vcorrection{-.7in}
\hcorrection{-.6in}

\newcommand{\C}[1]{{\cal#1}} 
\newcommand{\F}[1]{{\frak#1}}

\begin{document}

\addtolength{\footskip}{.3in}

\title{The Search for a New Equivalence Principle}
\author{Lucian Miti Ionescu}
\address{Department of Mathematics, Illinois State University, IL 61790-4520}
\email{lmiones@@ilstu.edu}


\begin{abstract}
The new emerging quantum physics - quantum computing conceptual bridge,
mandates a ``grand unification'' of space-time-matter {\em and} 
quantum information (all quantized),
with deep implications for science in general.

The major physics revolution in our understanding of the universe are
reviewed, pointing to the leitmotiv regarding their origin: 
a new fundamental principle,
usually expressible as an {\em equivalence principle}, 
captured the idea which provided the breakthrough,
while most of the technical theoretical tools were already in place.

Modern developments in quantum field theory
in relation to quantum computing, and in cosmology,
especially in connection with the black hole radiation laws,
represent supporting evidence for the existence of such an 
equivalence principle extending 
Einstein's well known equivalence principle $E=mc^2$.

The present article describes the nature of this ``missing'' equivalence principle
at a philosophical level, explaining the reasons for 
complexifying the energy-momentum tensor to include entropy,
and finally {\em unify space and time} by a generalized version of
the {\em Wick rotation} ``trick'':
$$\C{E}=ic\C{P},$$
an ``external supersymmetry'' which trades parallel computing and
sequential computing coordinates.

The usual {\em Feynman Path Integral} algebraic formalism$^{23,24}$ 
already has a place for this,
when interpreting entropy (information) as a measure of symmetry:
$$K(A,B)=\int\limits_{\Gamma\in Hom(A,B)} \C{D}\Gamma \ e^{H+iS(\Gamma)/\hbar}.$$
Additional details regarding the physical implementation$^{26}$ 
and the corresponding appropriate mathematical methods may be found in $^{14}$.
\end{abstract}

\maketitle
\tableofcontents

\section{Introduction}\label{C:VIRequest-UP}
Major breakthroughs in physics are based on a change in perspective 
regarding already existing mathematical models,
due to new {\em unifying principles}.

I will try to backup this statement and revisit some folklore fundamental questions and theoretical difficulties (``paradoxes'') 
which in the author's opinion should be solved 
as a result of a conceptual unification steaming from a new fundamental principle. 
At this stage (``proposal level'') we are able to list what seem to be 
the major pieces of a puzzle: a theory including the benefits of, 
and built with the technology of the present quantum theories 
(Quantum Mechanics/Quantum Field Theory, Quantum Gravity  etc.) 
while resolving the ``conceptual tension'' of the measurement paradox.
In my opinion, we do not always need express contradictions 
between experiment and theory.
Especially since nowadays, what theory predicts, 
say string theory for instance, 
may lie safely outside the experimental range.

The understanding may come from a new way of 
looking at the same ``technical tools'' 
(e.g. Special Relativity - see \S \ref{S:spec-rel}).

\section{Models, models, models!}\label{S:Models}
Recall that we model reality and we do not know what reality {\em is}; 
many books have been written on the subject, 
so I will only mention a few relevant names:
Kant, Mach etc. and revisit briefly a few relevant aspects.

\subsection{What do we mean by ``What is {\em time}?''?}
Implicitly we refer to a concept within a theory (framework/context etc.) 
which usually belongs to a specific community or person's knowledge, 
linked via a tag like Newton, Einstein, Heisenberg, Feynman etc.. 
Or, when asking ``What is an Electron?'', 
the answer ... depends on some ``hidden variables'': the theory we
implicitly have in mind when we ask that question.
``Electron'' may refer to the corresponding particle in Lorentz's theory, 
or the de Broglie's wave, Dirac's spinor etc. 
Even worse still, it can be quite misleading when ``explaining'' quantum mechanics, 
and in the same statement making use of the term ``electron'' to refer to 
the quantum description and then to ponder in classical terms about it ...

In this sense, there are many meanings behind what we call ``time'' or ``space'' etc., 
within various theories; so one has to be careful about the implied context.

\subsection{``Is it a particle or a wave?''}
The ``electron'' for instance, 
is very well modeled as a particle by a few theories, 
when it comes to a certain range of experiments, 
yet there is a need for other theories modeling the ``electron'' 
as a wave because of another class of experiments ... 
Overall quantum theory has a unified explanation for ``all'' experiments 
(of a certain kind, of course) and the {\em Complementarity Principle}
may be thought of as a ``Two classical charts atlas of Quantum Mechanics''.

\subsection{Interpretations of quantum mechanics}\label{S:int-QM}
Why do we need to ``interpret the result'' of a quantum mechanics computation 
in classical terms?

Classical mechanics is contained within quantum mechanics ($^4$, p.12), 
and it is not just a ``limit'' ({\em Correspondence principle}). 
Indeed, the measurement process involves a quantum probe
(microsystem, elementary particle etc.),
{\em interacting} with the measuring apparatus (usually a macrosystem) 
and the result of the experiment itself is modeled,
or at least used by the experimental physicist (or processed by some software!),
in classical terms. {\em We} acknowledge only macroscopical events: 
dots on a screen, beeps in a counter, bobbles in a chamber etc.. 
Even a Stern-Gerlach experiment (i.e. involving ``internal states'') 
involves the interaction of a quantum probe (the electron) 
with a magnetic field (macrosystem!) AND a detector: beeps on 2 counters,
providing the input to a classical gate/computation.
So, in a way Quantum Mechanics is a phenomenological theory! 
(beyond the Kantian statement of the type ``we only model phenomena ...''),
and the reduction from a quantum computation to a classical computation
mandates the collapse of the wave function: content lost in ``translation''.

\subsection{The main ``lesson''}\label{S:main-lesson}
from above is that there are implicit {\em channels of information} which are present, 
yet probably not correctly (or completely) modeled within the corresponding theory! 
The role of {\em observer} in classical physics is that of ``user'',
while in quantum physics is that of a ``quantum programmer''.
A crucial objective in classical physics is to have a {\em unique} description
(``standard operating system'' for the only hardware available)
independent of observer (covariance; classical heritage).
This is no longer tenable in quantum mechanics: ``results'' depend 
not only on ``what'' we observe (the quantum hardware the experimenter
physicist prepares) but also on ``how'' we observe, 
which in turn depends on what do you intend to do with ``the result''
(the quantum software used to model the quantum computation).
Nevertheless we are still looking for a 
``standard'' in these proliferation of quantum ``hardware-software'' business.

A unifying point of view, as a ``slogan'', 
if one has in mind the unification alluded to above (math-physics-computer science), 
is that ``All is quantum computing'' (see also ``Feynman processor'' $^1$ etc.),
i.e. any interaction, whether system-system (Einstein:
``I like (!) to think I don't have to look at the Moon for it to exist''), 
system-observer (quantum phenomenon), observer-observer (genuine communication,
predominantly classical!) 
are of the same kind.

\section{Three ... revolutions: returning to principles!}
Let us consider Newton's simplifying picture of Kepler's Laws 
as a start for scientific modeling of (mechanical) phenomena.
\footnote{Or ... is it ``a culminating point of the scientific revolution of 
the seventeenth century''? $^2$, p.425}

\subsection{First revolution: Special Relativity}\label{S:spec-rel}
{\em Special Relativity} gave a new look at the technical tools already available 
at that time: Minkowski space, Lorentz contraction, conformal invariance of 
Maxwell's equations etc. 
Yet the {\em conceptual} break-through consists in ``understanding'' their 
``hidden'' meaning: 
the unification of space and time.
Technically speaking, this was already done by Lorentz and Poincare - 
see $^3$, p.25 - but ... ``What is it that we are doing?'' was probably
the main question of the day. 
The unification was derived in an ``axiomatic'' manner 
from the fundamental principle $c=constant$,
which mathematically corresponds to a constant Lorentz metric (conformal class).
A probably more important fundamental principle is the 
equivalence between mass and energy:
$$Principle\ I:\quad E=mc^2.$$
A ``simple equation'' yet with huge implications.

\subsection{The second revolution: General Relativity (GR) or Quantum Mechanics (QM)?}
In the author's opinion, QM is {\em The} Revolution, 
changing the way physics is done (see \ref{S:main-lesson}).
{\em General Relativity} is a ``jewel'' amongst mathematical-physics theories, 
again starting from a fundamental principle, 
the equivalence between accelerations, gravitational or not,
or in terms of masses: inertial or gravitational):
$$Principle\ II:\quad m_g=m_a$$
General Relativity ``upgrades'' the Newtonian geometro-dynamic description 
``force of some kind=centripetal force'':
$$Force \ =\ Mass\ \times Acceleration$$
to a pure geometric description 
(space and time were already merged in special relativity)
``matter tensor $\sim$ geometry tensor'':
$$Matter\_Tensor\ =\ \kappa\quad Einstein\_Tensor.$$
Beyond the new ``technical tools'', e.g. semi-Riemannian spaces, Ricci curvature etc.,
this amounts to passing from a description of dynamics as 
``{\em curved} motion in {\em flat} (universal) space'' to 
``{\em flat} motion (geodesics) in {\em curved} space(-time)''.

In other words, taking a phenomenon (gravitational force for instance) 
from  the left hand side (LHS) of Newton's principle and 
incorporating it into the RHS as Einstein's tensor,
which is essentially the average curvature ($\kappa$ denotes the gravitational constant).
The ``trick'' proliferated: 
then came Kaluza-Klein, attempting the same maneuver with the electro-magnetic force. 
It did not work as well, since ``internal degrees of freedom'' 
could not be well accommodated as external degrees of freedom 
(i.e. dimensions of space-time). 
The alternative was to build degrees of freedom outside the ``obvious'' ones, 
leading to {\em Gauge Theory} (e.g. Yang-Mills theory etc.). 
Meanwhile the mathematics ``technology'' advanced and 
{\em String Theory} is capable of such feats, 
introducing ``real'' dimensions (for a grand total of 11? or 21? or ... etc.). 
Some of them, of course, need to be ``hidden'' from every-day ``access'' by compactification,
declaring them small enough not to contradict our experience.
But they are just ... another model for space-time!

In the ``phenomenological camp'' the opposite tendency may be noticed 
(in the spirit of quantum mechanics; see \ref{S:int-QM}): 
let the degrees of freedom (and states) be ``internal'' (abstract) ... 
and {\em Chiral algebras}, {\em Vertex Operator Algebras} etc. appeared!

So, where is the third revolution?

\subsection{Space-Time: Is ``motion'' possible?}
We do not need Zeno's paradox (see $^4$, p.56)
to claim that motion is not possible ($^5$, p.14)
\footnote{Zeno's {\em Arrow} paradox seams to urge for Lorentzian contraction at least.}. 
Of course, we have to specify in which theory:
in quantum mechanics, since otherwise classical mechanics 
deals great with motion/continuous evolution/dynamics (Poisson manifolds etc.), 
and we've learned not to talk about what reality {\em is}, 
but only modestly about our best model about it.

In quantum mechanics there are ``states'' and ``transitions'', 
as in a sort of a ``complexified'' Markov process, 
where, amazingly, the possibility of having a result in two ways 
may cancel each other's contribution (``indecision''!?),
rather then build up the probability! 
To model mathematically this feature,
we choose {\em superposition} and {\em interference}, 
implemented as a linear theory over complex numbers.

The incompatibility between knowing the position {\em and} 
the momentum at the same time, 
for the same direction (Heisenberg's uncertainty principle), 
conceptually refutes classical trajectories altogether,
but still refers to classical concepts!. 

If we insist in adopting QM to investigate the motion process
and still have a classical understanding of what the electron ``does'' 
in a two slit experiment we have to conclude that ``it'' goes 
through both holes simultaneously! 
This is in the ``best approximating'' classical statement for the quantum occasion ...
So, ``Is {\em motion} possible?'' Well ... the answer is theory (and author) dependent.

\subsection{What is an ``Event''?}
The differences in the approaches of modeling reality
in {\em Classical Mechanics}, General Relativity, Quantum Mechanics, Quantum Field Theory (QFT) 
start with the concept of ``event''$^7$.

For Newton the ``event'' is a ``particle'',
(i.e. {\em existence} of matter), ``somewhere in time''; 
these three concepts, existence, space and time, are ``absolute'', 
i.e. independent of the observer and of each other. 

For Einstein, ``existence'' is still ``absolute'', 
although the ``event'' occurs in a (partially) unified space-time,
yet still ``absolute'', even after the advent of GR. 
After the QM lesson, 
we should agree that what we model are {\em correlations}: 
$A$ interacting with $B$ produces $C$,
for example an {\em electron} in a {\em magnetic field} yielding a 
{\em beep} on the up or a down particle counter; 
... and the observer? 
There is a missing aspect here
in a parallel between quantum and classical computation$^8$,
to be explained elsewhere.

To implement ``correlations''  one needs to define the ``states'' and ``transitions'' 
(e.g. using categories: objects and morphisms). 
There is usually a ``time-ordering'' issue here: 
states first,  then transitions ... 
This may be thought of as developing the theory starting from the ``free case'' 
(inertial reference frames and the free theory in the scattering method approach) 
and then adding ``interactions'' (all frames/scattering matrix etc). 
It is essentially the old Newton's goal (and Descartes' methodology)
of representing functions as power series (or breaking down the theory in simpler steps).

In perturbative QFT the series is indexed by Feynman graphs, Riemann surfaces etc., 
i.e. building the ``big processor'' out of ``microcomponents''.

\subsection{Quantum Field Theory (QFT)}
In QFT we have a continuum of  degrees of freedom (the values of the field) 
only because we strongly believe in a given space-time continuum.
This is essentially the heritage of Newton and Leibnitz,
preserved in the classical theory of Einstein.

Roughly speaking, QFT is an ``upgrade'' of QM as a 
complexified Markov Process,
where the complete graph being {\em represented} (the transition matrix),
is replaced with a class of graphs and the complex numbers as coefficients 
are replaced with operators (propagators).

Feynman's path integral picture introduced what we will call 
{\em The Automaton Picture}: states and transitions, 
whether these are paths in space-time (external DOF) 
or transitions in internal space (IDOF).
This is a ``popular picture'' amongst physicists,
and at the same time the most powerful quantization method
(``Feynman brought QFT to the masses'' ($^9$, p.41).

We ultimately look for transition amplitudes of an interaction 
in the context of a framework based on the free case,
which is classical in essence since we know how many particles go in, 
and what comes out, in classical terms.
The amplitude is the sum of the amplitudes for all possible ``scenarios'';
the {\em correlation function} is a  sum over Feynman diagrams or possible histories. 
This is  a {\em basis} in the {\em transition space} (space of all ``paths''). 

The ``problem'' is, 
that if we believe ``motion'' is possible in a space-time continuum, 
then we end up with too many ``paths''!
This entails divergent integrals etc..

Physicists have learned quickly how not to step in quick send, 
while mathematicians had a hard time building the bridge over the 
``swamp of infinities'' 
(constant/variable, infrared/ultraviolet, important/neglectable etc.),
again due to the cherished inheritance of Newton and Leibnitz:
the {\em analysis} (doubt now ``hard analysis'').
It is time to acknowledge that the paths themselves of a given model are irrelevant,
forming the huge loop/path space of a Newtonian-Einsteinian version
of a configuration space of classical-mechanistic events.
Relevant is the homology/homotopy of the mode of interaction,
viewed as a {\em network processing quantum information},
as it will be explained in detail elsewhere $^{22}$
(see also $^{23}$).

\subsection{External/Internal Degrees of Freedom: {\em The Automaton Picture}}
The natural way to ``solve'' the problem of too many paths
is to realize that all we need is a reasonable category of ``paths'' (transitions) 
and an action allowing to build a representation of this ``Feynman category'' $^{23,24}$
with suited coefficients corresponding to 
the internal degrees of freedom had in mind.

By now it appears that gravity is an organizational principle 
within the space-time description (GR), rather than an exchange interaction. 
Trying to push the beautiful particle-field picture (gauge theory) 
from scalar and vector fields to spin 2 tensor fields and incorporate gravity
at all costs, could be the ``take a bigger hammer'' approach
(like from string theory to ... M-theory too) to ``crash the nutshell'', 
approach which looked so repelling to some (notably Grothendieck). 
It worked with Fermat's Theorem, though, but, ``What's taking so long?'' $^{10}$.

Alternatively, we could try to implement gravity as a pairing between the Feynman category
and the coefficient category.
The Feynman category captures the causality,
since there is NO universal time at the micro scale, 
and we have to deal with the time ordered products and operator product expansions
of QFT and renormalization.
The ``coefficient category'' captures 
the macro-behavior (see \S \ref{S:int-QM}) 
in an adjunction which trades additional external degrees of freedom 
(e.g. applying the homology differential, 
i.e. insertion of an edge$^{11}$) 
for additional internal degrees of freedom.
In this article, we focus on the ideas and design of the theory,
and therefore the technical details$^{12,13,15}$,
should not clutter the picture at this point!

This should be done in conjunction with a 
model for the information flow (see \S\ref{S:main-lesson}), 
since there are several macrosystems involved, 
and an experiment, like a quantum computation, 
involves classical read/write operations subject to classical 
logic/laws (see $^{14}$ for additional details).

No matter what the specific implementation will be 
(e.g. using graphs, networks, categories etc.), 
it will capture the idea of {\em automaton}:
{\em states} and {\em transitions}, 
e.g. the {\em cellular automata}
of$^{16,17}$). 
Yet the implementation will be written in one's favorite 
object-and-relations oriented high level language,
in an ``author dependent'' fashion.

\subsection{Is there a ``time'', after all?}
Indeed ``time'' is THE delicate concept; 
or rather a plethora of interconnected concepts! 
We all like to ponder on the fundamental questions, 
trying to find new ways ... (see ``Time's Up, Einstein'', 
by Josh McHugh, Wired 06/2005, p.122).
It was the analysis of what time is, 
that led Einstein to a clear picture 
unifying Newton's universal space with his universal time. 
Even at that stage,
one could ponder on a hidden assumption Einstein implicitly made: 
transitivity of synchronization.
It can indeed fail in GR, 
if there is no local time,
i.e. if the orthogonal distribution 
to the Killing vector field is not integrable$^{18}$. 
Instead of spending \$200,000 on a ``Michelson-Morley experiment'' 
trying to reintroduce the ``ether'' $^{19}$,
one might rather test the above mentioned possibility, 
which definitely holds true at {\em some} level of accuracy.

But since we aim at a deeper model,
beyond the Standard Model or String Theory, 
where ``events'' are ``pure correlations'', the above issues are secondary.
One lesson learned from Special Relativity is that there is a causal cone; 
events can be spatial separated (no causal correlation possible 
- we are not talking about entanglement yet ...),
or if causally correlated, 
than they must be time-separated. 
Yes, a ``proper time'', is a different concept,
representing ``continuity of existence'' and 
rather playing the role of a local parameter,
as opposed to the experimenter's global ``laboratory time'' in quantum mechanics.

So, what we need is a {\em Causal Structure} and 
that is precisely what a {\em Feynman Category} provides!
$$Feynman\ Category\quad \Longrightarrow \quad Causal\ Structure.$$
If a causality structure is given,
then to benefit from the present and past theories 
one has to deal with embedding it in a classical 
$d=4$ dimensional manifold (or is it $d=11, 21$?), 
as some ``background space''; 
or at least,  after representing it in one way or another 
(e.g. decorating punctures on Riemann surfaces with operators, 
or implementing algebraically as vertex operator algebras etc.) 
one has to come up with an {\em Operator Product Expansion} (OPE) 
as a much more complicated issue that the usual 
1-parameter group of unitary transformations
capturing the dynamical evolution of a mechanical system
as the ``time flow''.

Then, what is left of the idea of 4-coordinates as a 
``... starting point of the mathematical treatment'' ($^3$,p.24)? 
First of all, one should postpone the ``mathematical treatment'' 
until the ``design'' of the theory at a conceptual level is complete
or at least satisfactory: the {\em application interface} 
as a set of ``implementation specifications'' of the physics model.
Then let the implementation specialists (``math-programmers'') 
to chose the appropriate tool box to implement the theory ... 
But this is another story! we would not have had QED a few decades ago, right?
It had to be done fast, no time to wait for mathematicians to be 
pleased with a ``rigorous'', i.e. mathematical, implementation!  
What I am advocating here is to glance at the methodology of computer science,
and {\em design} the theory with an ``author independent'' and ``user friendly''
{\em interface} between mathematical and physics models.

On the other hand, there are some holistic questions.
There are 3-pairs of non-commuting observables 
representing external degrees of freedom ($q_1$, $p_1$, etc.). 
Why are there three dimensions? 
Why are there three generations of elementary particles? etc..
These could be questions allowing to tell theories apart, 
but we feel there is much more to these questions than it meets the eye.
They should be addressed as part of the quest for quantum gravity.

\section{A New Equivalence Principle}
Returning to General Relativity (GR), 
its importance still lies in the 
conceptual unification between space-time and matter.
At a more technical level,
perhaps the most important consequence
beyond expansion of universe and Hubble's constant, 
is the {\em concept} of black hole. 
The unification of GR and quantum theory was initiated by 
S. Hawking as an extension of GR incorporating the black hole radiation. 
Since then, three laws have been identified (see $^{20}$, p.92).
In view of the above unification and the main characteristics of a black hole,
namely the presence of an event horizon, 
we claim that the black hole of GR plays the role of the elementary particle 
from QFT: a ``black-box'' with internal degrees of freedom (DOF).
Therefore the black-hole radiation laws are much more fundamental
than presently acknowledged.

The first law relates {\em temperature}, 
as a measure of energy per DOF, 
with {\em acceleration} as a measure of the interaction (Newton's sense):
$$Unruh's\ Law:\qquad 	Temperature/\hbar=\ Acceleration/c.$$
It expresses a principle, 
therefore in the simplest (physicist favorite) way, as a linear equation. 
Together with {\em Einstein's Equivalence Principle}, 
it suggests that there is an energy distribution for 
the 2-point gravitational correlation function 
(in some quantum discrete picture).

The second law:
$$Bekstein's\ Law:\qquad \hbar\ Entropy\ =\ \frac1{\kappa}\ Area/(8\pi),$$
relates {\em entropy}, as a measure of the information needed 
to completely specify a state (the ``quantum memory size'')  and {\em area}, 
which in a discrete geometric model should be thought of 
as a measure of the possible In/Out interactions (``quantum channel capacity''). 
Beyond the ``global statement'', adequate for stating an equivalence principle, 
there should be here a ``local/discrete'' version (Stokes Theorem at work). 

It is reassuring to find out that Lee Smolin mentions implicitly 
such a ``would-be'' principle:
``one pixel corresponds to four Plank areas''$^{20}$ (p.90),
although it could rather be stated as
``one {\em interaction qubit} corresponds to four Plank areas''.

Later on (p.102), he derives some conceptual implications 
which are evaluated as not admissible, 
{\em IF} there is no theory to back them up 
(we have learned a lot from the old story: ``Euclid's Parallels, axiom or not?'';
let's derive the ``unbelievable'' consequences first,
then decide how to build the theory!).

Finally the third law relating temperature and mass, 
but in an opposite way as the first law, is:
$$Hawking's\ Law:\qquad	Temperature\ =\ k / Mass,$$
or alternatively:
$$Mass\ =\ k\beta$$ 
(with an eye on the entropy: Boltzmann's correspondence etc.).

It refers to the radiation capability of a black hole 
(``density of I/O-interactions''), 
rather then its energy distribution per DOF.

The situation is reminiscent of Newton's position 
when simplifying Kepler's laws ... 
so let's look for a new unifying principle, 
generalizing Einstein's Equivalence principle!

\subsection{``Mind versus matter''}
Recall that $E=mc^2$ (Principle I), in a sense, 
unifies energy and matter. 

Quantizing energy and matter (Planck, Einstein, Bohr, de Broglie etc.):
$$Principle\ III:\qquad 	Energy\ =\ \hbar \ Frequency\qquad (E=\hbar \omega)$$
should correspond to {\em quantizing quantum information}.
Then, since energy and matter determine space-time in GR,
space-time should be equally quantized.
Intuitively, quantum information (qubits) should be ``stored'' in 
quantum memory and processed by quantum gates and circuits.
This is precisely the role of space-time as a causal structure.
Loosely speaking, Feynman diagrams process quantum information
residing on the boundary$^{22,23,24}$. 

The new unifying {\em Equivalence Principle} will be labeled 
``Mind versus Matter'' to convey its broad scope. 
It states a correspondence between matter-energy and space-time-information, 
both quantized, and therefore discrete:
$$\text{\em New Equivalence Principle IV:}
\qquad qbit \leftrightarrow \hbar\qquad (S(qbit)=\hbar).$$
The left hand side represents the quanta of information (entropy),
and mathematically corresponds to a superposition of ``Yes'' and ``No''
with complex coefficients (probability amplitudes):
elements of a 3D sphere $S^3$.
We prefer to identify qubits with $SU(2)$ rather than with quaternions,
which exhibits the direct connection with {\em symmetry},
with its measure, the {\em entropy} (see$^{14}$, pp.104, 134):
$$Entropy\ <-> \ Symmetry:\quad H=-\ln Aut(\Gamma).$$
Here $\Gamma$ represents the state space with its symmetries $Aut(\Gamma)$.

The right hand side is the unit of action, the Plank's constant
as a suggestive symbol of quantum physics and quantization.

So far we aim not just to unify
the ``observer'' and ``observed'' of quantum physics, 
and resolving the ``measurement paradox'', 
but because the usual current resolutions of this paradox 
involve at some point the conscience,
we also provide a possible interface between the ``safe'' science and 
the other ``believe-it-or-not'' areas of investigation
(direct interactions between matter and mind).

The idea is that a transfer or fluctuation of a unit of energy 
should correspond to a quantum bit of information 
An additional DOF (E/I) (internal, i.e. type of particle, 
or external (!), i.e. space-time ``location''$^{14}$) 
changes the partition function describing the distribution of amplitudes 
of probabilities in a way similar 
to a black hole ``leaking'' a qubit of information. 
The theory should naturally incorporate the black hole laws 
in the context of GR transmuted 
from its natural habitat (manifolds with a metric/Lagrangian) 
to the realm of {\em Feynman Processes}
(representations of Feynman Categories: string/M-theory rephrased 
as background free theories, with a mass generation mechanism
upgrading the Higgs breaking of symmetry).

At the more technical level of the Feynman Path Integral
formalism, a conceptual ``merger'' between energy and entropy
can be achieved with the price of {\em complexifying the action}$^{14}, p.224$:
$$K(A,B)=\int\limits_{\Gamma\in Hom(A,B)}\frac{e^{iS(\Gamma)/\hbar}}{|Aut(\Gamma)|}
=\int_A^B \C{D}\Gamma \ e^{H+iS(\Gamma)/\hbar}.$$
This is mandated by chiral theories anyway, 
where the Lagrangian contains complex coupling constants
to account for charge-parity symmetry violations in 
particle physics interactions $^{25}$.
Another reason for complexifying the action is the conclusion ($^{12}$, p.225)
that it is {\em not} space in need of extra dimensions (11, 21 etc.
as in ``classical'' String Theory),
but rather {\em time} (st-symmetry in conformal field theory, Wick rotation
to switch from Minkovskian to Euclidean QFT etc.).
The new reason for enlarging the usual 1-dimensional time flow of mechanics
to a 3-dimensional flow, is the interpretation of interactions as 
communications of quantum information (qubits),
towards a representation theory of a {\em non-abelian time}: $SU(2)$,
instead of the classical 1-parameter {\em abelian time}: $\F{R}$.
This will indeed {\em unify space and time} (ST), 
as playing the {\em dual role of 
parallel and sequential quantum computation coordinate system}
in quantum networks admitting such a 
{\space-time coordinate system} (see$^{26}$ for additional details).
The alternative for the mathematical implementation 
triples the number of time dimensions for a total of 3+3 ST-dimensions
or doubles the ST (2x(3+1).

At an elementary level (the physics interface),
the equivalence between energy and matter is expressible in the 
well-known simple manner: $E=mc^2$; or rather $E=c^2 m$,
exhibiting the fundamental role of the ``speed'' of light,
as a proportionality constant:
$$E^2=c^2[p^2+(m_0c)^2].$$
To unify with entropy/information, we include an entropy term,
which renders an additional degree of symmetry of the above formula:
\begin{equation}\label{E:NEP}
E^2+\hbar^2H^2=c^2(p^2+p_0^2).
\end{equation}
The formula now reflects the {\em IE-duality} of the
Quantum Dot Resolution (QDR), between external and internal DOFs$^{14}$.

The further unification of matter-energy and space-time
in Einstein's GR can be suggested by the ``simple'' formula:
$$Geometry\ \sim \ Energy-Momentum:\quad G=\chi T.$$
In a similar way, we simplify our Equation \ref{E:NEP}
by complexifying the framework, 
as needed by the many other reason stated above.
Introduce the 
$$Complex\ Energy\ and\ Momentum:\quad \C{E}=E+ik_BT H, \quad
\C{P}=p+icm_0,$$
where $k_B$ denotes Boltzmann's constant.
The entropy $H$, or rather the extensive quantity $TH$ of p.104,
is related to a potential function $Q$$^{14}$, p.78:
$$H(\partial \Gamma)=Q(\Gamma)\quad (TH=dQ),$$
which may be related to the quantum potential of Bohmian mechanics
$^{14}$, p.192. 

Then the above equality of moduli should be the shadow of a complex
energy-momentum (tensor) equivalence, 
which expresses the equivalence between space and time coordinates 
at the level of the corresponding canonically conjugate variables:
$${\bf Generalized\ Wick\ Rotation:\quad \C{E}=ci\C{P}}.$$
It is consistent with Plank quantization $E=h\nu$
and de Broglie wave-particle duality $cp=h\nu$.
Moreover, the complex energy-momentum tensor
$$\C{E}=E+iQ, \quad \C{P}=p_e+ip_i$$
provides a unified framework for the energy-momentum flow
of external DOFs ($p=p_e$) {\em and}
quantum information flow of internal DOFs ($p_i=cm_0$).

A detailed implementation is scheduled to appear in 2008$^{26}$.

\subsection{Are black holes prototypical?}
Again it is reassuring that the idea of the above unifying 
Equivalence Principle, 
in a form hinting towards the relation between space-time and information, 
is present in the remarkable book$^{20}$ (p.101): 
``There is something incomplete about a law which asserts a balance 
or an exchange between two very dissimilar things.''. 
Paying too much attention to its ``draw backs'' is not always a good idea 
(loc. cit. p.102). 
Yes, if one would just claim $E=mc^2$, would not be enough ... 
But again, a theory starts with an idea, a new principle 
(1\% of inspiration), 
and then one designs the theory top-down (99\% of the perspiration - T. A. Edison).

So, Lee Smolin is talking about a balance between ``atoms'' and ``geometry'', 
which seems to him an unacceptable ``mix'' within the current theories (true!).
But only in gauge theories on {\em manifolds} 
(or fixed lattices approximating them),
there is a clear cut distinction between 
external DOF, modeled as a space-time, 
and internal DOF implementing the type of particle 
as a representation of a gauge group,
to be ``married'' as a principle bundle etc..
Moreover, a distinction between ``atoms'' and ``geometry'' 
still leads back to an ``absolute space-time'' point of view. 

This is no longer true in a {\em Feynman-Ionescu Theory} 
(FPI adapted to {\em Feynman Processes} as representations of 
{\em Feynman Categories} modeling causality and quantum information flow) 
where an insertion of a new graph should be thought of as ``adding geometry'' 
(and also as a change of scale!, i.e. {\em resolving} 
additional DOFs in the QDR multi-resolution analysis
extending the fixed lattice point of view,
e.g. LGT: Lattice Gauge Theory).
Then, under the functorial adjunction (IE-duality), 
new internal DOFs are introduced: quantum information/qubits.

Now my ``bet'' regarding the two profound questions from$^{20}$, p.102, is:

\quad (A) Yes, there is an ``atomic structure'' of the geometry of space-time,
i.e. it leads to a better model (see \ref{S:Models}),
e.g. PROPs or LQG's ``grains of space-time''$^{23,24}$
and also $^{21}$.

Our unifying {\em New Equivalence Principle} generalizes in a sense 
the idea behind Bekstein's Law. 
Indeed, in a discrete {\em Feynman Category} model, 
``area'' corresponds to the number of interactions,
which from Unruh's law, ``carry'' a certain energy.
Roughly speaking a ``space-time event'' $A\to B$ 
has a double role of both {\em interaction channel} and {\em information channel}.

\quad B) Yes, the {\em Digital World Theory}$^{14}$ 
incorporating the theory of information 
(Shannon, quantum computing etc.) on top of a {\em Feynman Theory}, 
will have as natural consequences the black hole radiation laws,
but in a different disguise (discrete of finite type).

How to switch from black holes, 
thought of as ``prototypical'' when it comes to ``global'' quantum aspects, 
to the general case, say in terms of Feynman graphs? 
It is too soon for ``technicalities''! ... at least here,
in this essay$^{26}$.

\section{Conclusions}\label{S:Conclusions}
Let us review the main ``evolutionary steps'' of 
the fundamental concepts:

\vspace{.2in}
\begin{center}
\begin{tabular}{|l|c|c|c|c|} \hline
Newton & Space & Time & Particle xor wave & N.A.\\ \hline
Einstein & \multicolumn{2}{c|}{Space-Time} & Particle xor wave & Observer\\ \hline
Heisenberg & Space & Time & \multicolumn{2}{c|}{Particle/Wave \& Observer}\\ \hline
Dirac & Space-Time & \multicolumn{3}{c|}{Particle/Wave \& Observer} \\ \hline
Feynman & \multicolumn{4}{c|}{Path Integral Quantization}\\ \hline
Math.Ph.-Folklore &	
\multicolumn{4}{c|}{Representations of Feynman Categories}\\ \hline
{\em The DWT v 2.0} & 
\multicolumn{4}{c|}{Representations of Causal Structures with IE-duality:} \\ 
\qquad $\C{E}=ic\C{P}$ & \multicolumn{4}{c|}{{\em Hodge-de Rham Quantum Dot Resolution}.}\\
\hline
\end{tabular}
\end{center}

\vspace{.2in}
Here the ``extended'' {\em Causal Structure with IE-duality}
refers to the incorporation of 
the concepts of entropy and information processing $^{22}$,
in order to unify the classical interactions ``particle-particle'' 
and ``particle-observer'' modeled by {\em Quantum Theory}
with ``observer-observer'', 
i.e. genuine communications.
Besides the symmetry reasons, 
the author hopes that it would lead to a better understanding of ``reality'',
for example of the measurement paradox and of 
the ``final frontier'': the {\em Mind-Matter Interface}.

How to put together all the above ``design constraints'' in a coherent theory,
is another story$^{26}$.
Its interface is {\em The Virtual Institute}$^{29}$,
intended to stimulate the upbringing of 
{\em The Digital World Theory}$^{14}$:
$$\mbox{Reality\ {\em is}\ ``The\ Quantum\ Matrix''!}$$




$^1$ G. J. Milburn,
The Feynman Processor: Quantum Entanglement and the Computing Revolution,
Frontiers of Science (Perseus Books), 1998.

$^2$ Victor J. Katz,
A history of mathematics, 2nd ed., Addison Wesley Longman, Inc., 1998.

$^3$ V. S. Varadarajan,
Supersymmetry for mathematicians: an introduction, Courant Lectures in Mathematics, 11.

$^4$ L. D. Landau and E. M. Lifshitz, 
Quantum mechanics: non-relativistic theory, 
Course of Theoretical Physics, Vol.3.

$^5$ Stephen L. Adler,
Quantum Theory as an Emergent Phenomenon,
Institute for Advanced Study, Princeton, New Jersey, 2004.

$^6$ Bohmian mechanics, http://www.math.rutgers.edu/\~{}oldstein/quote.html

$^7$ L. M. Ionescu,
Remarks on quantum physics and non-commutative geometry,\\
math.HO/0006024, 2000.

$^8$ L. M. Ionescu, Projects,
Virtual Institute for Research in Quantum Entropy,
Space and Time, {\em www.VIRequest.com}.

$^9$ A. Zee, QFT in a nutshell, 2003.

$^{10}$ S. Battersby,
Are we nearly there yet?, New Scientist, 30 April 2005, p.30.

$^{11}$ L. M. Ionescu,
Perturbative quantum field theory and configuration space integrals,
hep-th/0307062.

$^{12}$ L.M. Ionescu,
Perturbative Quantum Field Theory and L -Algebras, 
Advances in Topological Quantum Field Theory, 
Proceedings of the NATO ARW on New Techniques in Topological Quantum Field Theory, 
editor J. Bryden, Kluwer Academic Publishers, 2004, p. 243-252.

$^{13}$ L. M. Ionescu,
Cohomology of Feynman graphs and perturbative quantum field theory,
{\em Focus on Quantum Field Theory}, {\bf Vol.1}, 2004,
O. Kovras (editor), NovaScience Publishers, Inc..

$^{14}$ L. M. Ionescu,
The Digital World Theory, ed. Olimp Press, ISBN: 973-7744-39-x, 2006.

$^{15}$ D. Fiorenza and L. M. Ionescu,
Grand configuration spaces, Feynman integrals and renormalization,
NSF grant proposal and working project,\\
http://www.virequest.com/ISUP/VI\_ISU-GP.html
2006.

$^{16}$ Stephen Wolfram, A new kind of science, 2002.

$^{17}$ D. Shiga,
``Cells are circuits, too'', WIRED, Issue 13.04 - April 2005,\\
http://www.wired.com/wired/archive/13.04/start.html?pg=4

$^{18}$ B. O'neil,
Semi-Riemannian geometry. With applictions to relativity, Pure and Applied Mathematics, 103.

$^{18}$ Marcus Chow, Catching the cosmic wind,
New Scientist, 2 April 2005, p.30.

$^{19}$ Lee Smolin, Three roads to quantum gravity, 2001.

$^{20}$ Lee Smolin,
Atoms of space and time, 
Scientific American, special edition, Dec. 2005, p.56-66.

$^{21}$ VIReQuest Projects,
http://www.virequest.com/VIReQuest\_Projects.htm

$^{22}$ L. M. Ionescu, What space and time really are, in preparation.

$^{23}$ L. M. Ionescu, The Feynman Legacy, math.QA/0701069.

$^{24}$ L. M. Ionescu,
From operads and PROPs to Feynman processes, 
math.QA/0701299, to appear in JPANTA, 2007.

$^{25}$ G. D. Coughlan, J. E. Dodd 
and B. M. Gripaios,
The Ideas of Particle Physics, 3rd ed., Cambridge University Press, 2006.

$^{26}$ L. M. Ionescu, What {\em space} and {\em time}
really are, in preparation, 2007.

$^{27}$ L. M. Ionescu, Q++ and a nonstandard model, 
in preparation, 2007.

$^{28}$ L. M. Ionescu, The Hodge-de Rham theory of the
Quantum Dot Resolution, to appear 2008.

$^{29}$ L. M. Ionescu, The Virtual Institute for Research in
Quantum Entropy, Space and Time, www.VIReQuest.com.


\end{document}